\documentclass{aa}  
\usepackage{graphicx}
\usepackage{txfonts}
\usepackage{booktabs}

\newcommand{\swift}{\textit{Swift} }
\newcommand{\agile}{\textit{AGILE} }

\begin{document}

\title{A photometric redshift of $z=1.8^{+0.4}_{-0.3}$ for the \agile GRB 080514B}

\author{
A.~Rossi\inst{1},
A.~de~Ugarte~Postigo\inst{2},
P.~Ferrero\inst{1},
D.~A.~Kann\inst{1},
S.~Klose\inst{1},
S.~Schulze\inst{1},
J.~~Greiner\inst{3},
P.~~Schady\inst{4},
R.~Filgas\inst{1},
E.~E.~Gonsalves\inst{1,5},
A.~K\"upc\"u~Yolda\c{s}\inst{3,6},
T.~Kr\"uhler\inst{3,7},
G.~Szokoly\inst{3,8},
A.~Yolda\c{s}\inst{3},
P.~M.~J.~Afonso\inst{3},
C.~Clemens\inst{3},
J.~S.~Bloom\inst{9},
D.~A.~Perley\inst{9},
J.~P.~U. Fynbo\inst{10},
A. J. Castro-Tirado\inst{11},
J.~Gorosabel\inst{11},
P. Kub\'anek\inst{11,18},
A.~C.~Updike\inst{12},
D.~H.~Hartmann\inst{12},
A.~Giuliani\inst{13},
S.~T.~Holland \inst{14},
L.~Hanlon\inst{15},
M. Bremer\inst{16},
\and
A. Garc\'{\i}a-Hern\'andez \inst{17}
}

\offprints{A. Rossi, rossi@tls-tautenburg.de}

\institute{Th\"uringer Landessternwarte Tautenburg, Sternwarte 5,
   D--07778 Tautenburg, Germany
\and
   European Southern Observatory, Alonso de C\'ordova 3107, Vitacura, Casilla
   19001, Santiago 19, Chile
\and
   Max-Planck-Institut f\"ur Extraterrestrische Physik, 
   Giessenbachstrasse, D--85748 Garching, Germany
\and
   Mullard Space Science Laboratory, University College London, Holmbury St.
   Mary, Dorking, Surrey RH5 6NT, England
\and
   Darthmouth College, Hanover, NH 03755, USA
\and   
   European Southern Observatory, Karl-Schwarzschild-Strasse 2, D-85748
   Garching, Germany
\and
   Universe Cluster, Technische Universit\"{a}t M\"{u}nchen, Boltzmannstra\ss e
   2, D-85748, Garching, Germany
\and
   Institute of Physics, E\"otv\"os University, P\'azm\'any P. s. 1/A, 1117 
   Budapest, Hungary
\and
   Department of Astronomy, University of California, Berkeley, CA 94720-3411,
   USA 
\and
   Dark Cosmology Centre, Niels Bohr Institute, University of Copenhagen, 
   Juliane Maries Vej 30, DK--2100 Copenhagen, Denmark
\and
   Instituto de Astrof\'{\i}sica de Andaluc\'{\i}a (IAA-CSIC), Apartado 
   de Correos 3.004, E--18080 Granada, Spain
\and
   Clemson University, Department of Physics and Astronomy, 
   Clemson, SC 29634-0978, USA
\and
   INAF/IASF-Milano, Via E. Bassini 15, I-20133 Milano, Italy
\and
   NASA Goddard Space Flight Center, Greenbelt, MD, USA
\and
   College of Engineering, Mathematical \& Physical Sciences, School of Physics
   Ucd Science Centre, Belfield, Dublin 4, Ireland
\and
   Institut de Radio Astronomie Millimetrique (IRAM), 300 rue de la 
   Piscine, F-38406 Saint-Martin d\' \rm Heres, France
\and
   Instituto de Astrof\'{\i}sica de Canarias (IAC), C/. Via L\'actea s/n, 
   E-38205 La Laguna (Tenerife), Spain
\and
   Universidad de Valencia, Edif. Institutos de Investigaci\'on 
   (GACE-ICMOL), Campus de Paterna, E-46980 Paterna, Spain.
   }

\date{}
 
\authorrunning{Rossi et al.}
\titlerunning{Optical observations of GRB 080514B}

\abstract
{}
{The AGILE gamma-ray burst GRB 080514B is the first burst with 
detected emission above 30 MeV and an optical afterglow.
However, no spectroscopic redshift for this burst is known.}
{We compiled ground-based photometric optical/NIR and millimeter data from several
observatories, including the multi-channel imager GROND, as well as ultraviolet \swift
UVOT and X-ray XRT observations. The spectral energy distribution of the
optical/NIR afterglow shows a sharp drop in the \swift UVOT UV filters that
can be utilized for the estimation of a redshift.}
{Fitting the SED from the \swift UVOT $uvw2$ band to the $H$ band,
 we estimate a photometric redshift of $z=1.8^{+0.4}_{-0.3}$, 
consistent with the pseudo redshift reported by
Pelangeon \& Atteia (2008) based on the gamma-ray data.}
{The afterglow properties of GRB 080514B do not differ from those
exhibited by the global sample of long bursts, supporting the view that afterglow 
properties are basically independent of prompt emission properties.}


\maketitle


\section{Introduction}

Gamma-Ray Bursts (GRBs) are the most luminous explosions in the Universe, 
with the bulk of the released energy emerging in the 0.1 to 1 MeV range (e.g.,
Kaneko et al. 2006; Preece et al. 2000). Indeed, most bursts have not been
observed at energies much above 1 MeV, where low photon counts and typically
small instrumental collecting areas hamper the gathering of data. For
example, the \it Burst And Transient Source Experiment \rm (BATSE; operating
from 25 keV to 2 MeV) aboard the \it Compton Gamma-Ray Observatory \rm
(CGRO) detected 2704 bursts during a nine-year lifetime (1991 to 2000),
while the COMPTEL telescope on CGRO, operating in the  0.8 MeV to 30 MeV
range, in the same time period observed only 44 events with high significance
(Hoover et al. 2005). The number of GRBs detected at even higher energies with
the EGRET TASC and spark chamber instruments is even lower (Dingus 1995; Kaneko
et al. 2008). 

\begin{figure}
\includegraphics[width=8.8cm]{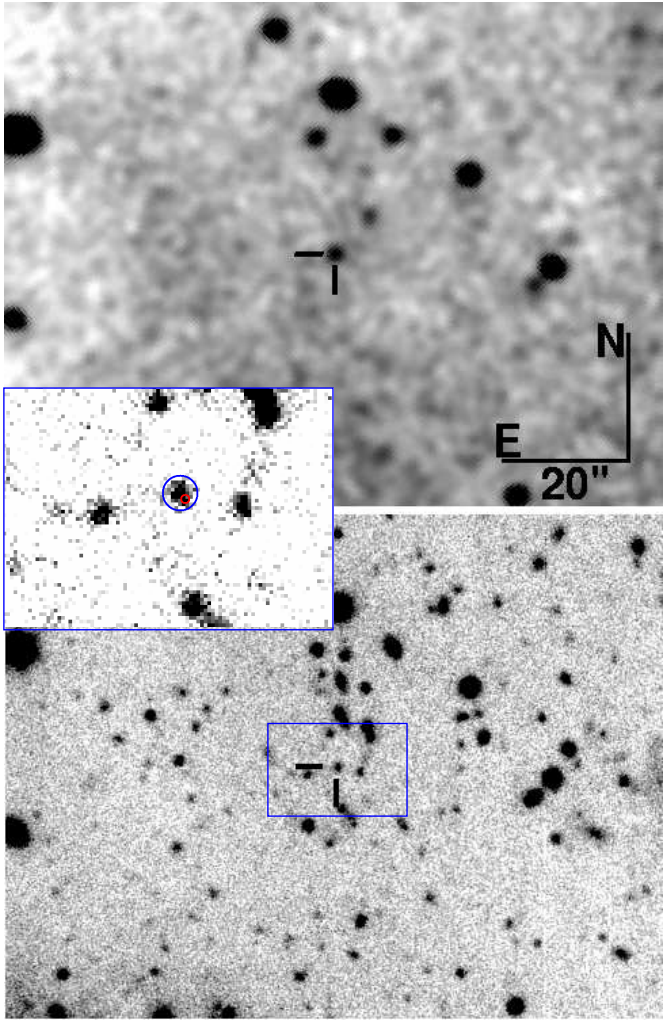}
\caption{\it Top: \rm IAC80 $I$-band discovery image of the optical 
afterglow of GRB 080514B (de Ugarte Postigo et al. 2008). The afterglow is highlighted. 
\it Bottom: \rm Keck $R$-band image obtained 24 days after the trigger. The underlying host galaxy 
is clearly detected. The zoom-in of the Keck image shows the host galaxy (encircled) and 
the GRB position (red circle).}
\label{fig:finding}
\end{figure}

To date no burst detected above 30 MeV has
an observed afterglow. The discovery of GRB 080514B by the Italian \agile gamma-ray 
satellite (Tavani et al. \cite{Tavani}) on May 14, 2008 at 09:55:56 UT
(Rapisarda et al. 2008) was therefore of particular interest. \agile carries
three instruments covering the energy range from 20 keV to 50 GeV and detected
GRB 080514B at energies well above 30 MeV (Giuliani et al. \cite{Giuliania},\cite{Giulianib}). 
GRB 080514B was a bright, multi-spiked event with a duration (T90) of 5.6 s 
in the energy range from 350 to 1000 keV, characterizing it as a long
burst.

The burst was also observed by Mars Odyssey, operating as part of the
Interplanetary Network (IPN) (Hurley et al. 2006), making it possible to
constrain the error box of the burst to about 100 arcmin$^2$ (Rapisarda et
al. 2008). This localization led to the discovery of its X-ray afterglow by 
the \swift satellite at coordinates R.A., Dec. (J2000) =  $21^{\rm h} 31^{\rm
m} 22 \fs62, +00^\circ 42\arcmin 30\farcs3$ with an uncertainty of 1\farcs6
(radius, 90\% confidence) at 0.43 days after the trigger (Page et al. 2008). 
Before the announcement of the X-ray afterglow position, however, the 
optical afterglow had already been discovered by our group by observing the 
complete IPN error box (de Ugarte Postigo et al. \cite{Postigoa}, 
\cite{Postigob}; fig.\ref{fig:finding}). Here we report our optical/near-infrared
follow-up observations of the afterglow of GRB 080514B starting 0.43 days
after the trigger, extending to late times of 24 days.

\begin{table}
\caption{Log of observations. The first column gives the mid-time
in days after the GRB. Vega magnitudes are 
not corrected for Galactic extinction. Also given are 3$\sigma$ upper limits.}
\begin{tabular}{rlccl}
\toprule
$t$ (days)& filter & instr./telesc. & exposure (s) & mag      \\
\midrule
0.430 &$UVW1$    & UVOT      & 284      & 20.45  $\pm$ 0.40 \\
0.432 & $U$	  & UVOT      & 142      & 19.75  $\pm$ 0.30 \\
0.434 & $B$	  & UVOT      & 142      & 21.00  $\pm$ 0.63 \\
0.438 &$UVW2$    & UVOT      & 568      & 21.47  $\pm$ 0.56 \\
0.443 & $V$	  & UVOT      & 142      & 20.88  $\pm$ 1.50 \\
0.446 &$UVM2$    & UVOT      & 413      & 21.97  $\pm$ 1.30 \\
0.488 &$UVW1$    & UVOT      & 419      & 20.17  $\pm$ 0.27 \\
0.492 & $U$	  & UVOT      & 209      & 19.91  $\pm$ 0.27 \\
0.494 & $B$	  & UVOT      & 209      & 20.40  $\pm$ 0.31 \\
0.501 &$UVW2$    & UVOT      & 838      & 21.78  $\pm$ 0.57 \\
0.507 & $V$	  & UVOT      & 209      & 19.74  $\pm$ 0.42 \\
0.512 &$UVM2$    & UVOT      & 616      & 20.63  $\pm$ 0.37 \\
0.555 &$UVW1$    & UVOT      & 415      & 20.88  $\pm$ 0.46 \\
0.559 & $U$	  & UVOT      & 207      & 20.09  $\pm$ 0.32 \\
0.561 & $B$	  & UVOT      & 207      & 21.62  $\pm$ 0.91 \\
0.567 &$UVM2$    & UVOT      & 791      & 22.09  $\pm$ 0.76 \\
0.640 & $R_C$    & Watcher   & 120x14   & 19.23  $\pm$ 0.47 \\ 
0.660 & $R_C$    & Watcher   & 120x15   & 19.89  $\pm$ 0.56 \\ 
0.727 & $I_C$    & IAC 80    & 3x300    & 20.26  $\pm$ 0.21 \\
0.743 & $I_C$    & IAC 80    & 3x300    & 20.59  $\pm$ 0.20 \\
0.761 & $I_C$    & IAC 80    & 3x300    & 20.16  $\pm$ 0.16 \\
0.774 & $I_C$    & IAC 80    & 3x300    & 20.03  $\pm$ 0.14 \\
0.907 & $g^\prime$   & GROND/2.2m& 3x1501   & 21.53  $\pm$ 0.04 \\
0.907 & $r^\prime$   & GROND/2.2m& 3x1501   & 21.16  $\pm$ 0.03 \\
0.907 & $i^\prime$   & GROND/2.2m& 3x1501   & 20.77  $\pm$ 0.08 \\ 
0.907 & $z^\prime$   & GROND/2.2m& 3x1501   & 20.43  $\pm$ 0.05 \\ 
0.907 & $J$    & GROND/2.2m& 3x1200   & 19.82  $\pm$ 0.03 \\ 
0.907 & $H$    & GROND/2.2m& 3x1200   & 19.10  $\pm$ 0.04 \\ 
0.907 & $K$    & GROND/2.2m& 2x1200   & $>$17.5          \\ 
1.021 & $J$    & NEWFIRM/KPNO&23x30x2 & 19.84  $\pm$ 0.14 \\ 
1.038 & $J$    & NEWFIRM/KPNO&15x30x2 & 20.06  $\pm$ 0.07 \\ 
1.763 & $R_C$  & NOT       & 1x300    & 22.31  $\pm$ 0.08 \\  
1.782 & $B$    & NOT       & 1x300    & 23.03  $\pm$ 0.13 \\
1.798 & $I_C$  & NOT       & 1x300    & 22.00  $\pm$ 0.10 \\
1.899 & $i'$  & GMOS/Gemini & 1x200    & 21.83  $\pm$ 0.06 \\
1.993 & $g^\prime$   & GROND/2.2m& 1x1501   & 22.74   $\pm$0.08  \\  
1.993 & $r^\prime$   & GROND/2.2m& 1x1501   & 22.38  $\pm$ 0.10 \\  
1.993 & $i^\prime$   & GROND/2.2m& 1x1501   & 21.78  $\pm$ 0.13 \\ 
1.993 & $z^\prime$   & GROND/2.2m& 1x1501   &  $>$21.6    \\   
1.993 & $J$    & GROND/2.2m&  1x1200  & $>$20.3 	 \\  
1.993 & $H$    & GROND/2.2m&  1x1200  & $>$19.1 	 \\   
1.993 & $K$    & GROND/2.2m&  1x1200  & $>$17.7 	 \\  
2.023  & $J$    & NEWFIRM/KPNO& 15x30x2& 20.95  $\pm$ 0.30 \\  
2.039  & $H$    & NEWFIRM/KPNO& 15x15x4& $>$20.3           \\
2.536 & White  & UVOT      & 5361     & 22.19  $\pm$ 0.17 \\ 
8.965 & $g^\prime$   & GROND/2.2m& 4x1501   & 24.05  $\pm$ 0.17 \\ 
8.965 & $r^\prime$   & GROND/2.2m& 4x1501   & 24.40  $\pm$ 0.25 \\ 
8.965 & $i^\prime$   & GROND/2.2m& 4x1501   & 23.35  $\pm$ 0.26 \\ 
8.965 & $z^\prime$   & GROND/2.2m& 4x1501   & 23.28  $\pm$ 0.24 \\  
8.965 & $J$    & GROND/2.2m&  4x1200   &  $>$21.9 \\ 
8.965 & $H$    & GROND/2.2m&  4x1200   &  $>$20.5 \\ 
8.965 & $K$    & GROND/2.2m&  3x1200   &  $>$18.4 \\ 
24.13  & $R_C$  & Keck      &    960    & 24.17  $\pm$ 0.33 \\
24.13  & $g^\prime$    & Keck &  1080      & 24.73  $\pm$ 0.34 \\
\bottomrule
\end{tabular}
\label{mags}
\end{table}
\section{Observations and data reduction}

\swift UVOT began observing the afterglow 0.43 days after the SuperAGILE/IPN
detection (Holland \cite{holland}), in the broad-band $V$, $B$, $U$, $uvw1$, 
$uvm2$ and $uvw2$ lenticular filters, covering the wavelength range between 
1600 {\AA} and 6000 {\AA} (Poole et al. 2008). A second set of observations were 
obtained in the $white$ band $\sim2.5$ days after the trigger, covering the 
wavelength range from 1600~{\AA} to 8000~{\AA}. Photometry was performed on the 
UVOT data using the standard \swift software tool {\it uvotmaghist} (version 1.0),
where source counts are extracted using a circular aperture with a radius of
3\arcsec, and the background was derived from a source-free region close to the 
target with a 15\arcsec \ radius. An aperture correction was applied in order to 
remain compatible with the effective area calibrations, based on 5\arcsec \ aperture 
photometry (Poole et al. 2008).

Ground-based follow-up observations were performed by our group using the $16''$ 
Watcher telescope in South Africa, the IAC80 telescope at Observatorio del
Teide, the MPG/ESO 2.2m telescope on La Silla equipped with GROND (Greiner et
al. 2007, 2008), the Nordic Optical Telescope on La Palma, the Kitt Peak 4m
telescope, the Gemini North 8m and the Keck 10m telescope on Mauna Kea, Hawaii (Table~\ref{mags}).
The data were analyzed using standard PSF photometry and aperture photometry for 
the host galaxy. The data given in the table 
supersede the magnitudes reported in the Gamma-ray burst Coordinate Network circulars 
(de Ugarte Postigo et al. \cite{Postigoa},\cite{Postigob}; Rossi et al. 2008a,b; 
Updike et al. 2008a,b; Malesani et al. \cite{Malesani}; Perley et al. \cite{Perley}).

Our dataset is completed by an observation at 86 GHz with the Plateau 
de Bure interferometer (Guilloteau et al. 1992) using the 5-antenna compact D 
configuration and performed 3.92 days after the burst. We did not detect any 
source at the afterglow position within a 3-sigma detection limit of 0.57 mJy.

Afterglow coordinates were derived from the GROND first epoch stacked $r'$-band
image, which has an absolute astrometric precision of about $0\farcs2$,
corresponding to the RMS accuracy of the USNO-B1 catalogue (Monet et al. 2003). 
The coordinates of the optical afterglow are R.A., Dec. (J2000)  =
$21^{\rm h}31^{\rm m}22.^{\rm s}69$, $+00^\circ42'28\farcs6$ (Galactic 
coordinates $l$, $b$ = $54.^\circ57\,,-34.^\circ49$). The Schlegel,
Finkbeiner, \& Davis (1998) extinction maps give $E(B-V)=0.06$ mag along
this line of sight through the Galaxy. Assuming a ratio of
visual-to-selective extinction of 3.1, this implies $A_V\approx$ 0.19 mag.

\swift performed a Target of Opportunity (ToO) observation of the
\emph{AGILE}/IPN error box 0.43 days after the trigger and found two new 
X-ray sources. The brightest of these was found to fade, identifying it as
the X-ray afterglow. 
X-ray data were obtained from the \swift data archive and the light curve from
the \swift\ light curve repository (Evans et al. 2007). To reduce the data, the
software package {\tt HeaSoft} 6.4 was used\footnote{see \tt
http://heasarc.gsfc.nasa.gov/docs/software/\\lheasoft/}, with the
calibration file version {\tt v011\footnote{see \tt
http://heasarc.gsfc.nasa.gov/docs/heasarc/caldb/\\caldb\_intro.html}}. Data
analysis was performed following the procedures described in Nousek et al. (2006). 
Spectral analysis was performed with the software package {\tt Xspec v12}, using 
the elemental abundance templates of the Galactic interstellar medium given by 
Wilms et al. (2000). Because \emph{Swift} did not begin observations until 0.43 days after 
the trigger, the quality of the spectrum and of the light curve of the X-ray
afterglow suffers from a low count rate and data gaps due to \emph{Swift}'s orbit.

\section{Results}

\subsection{The X-ray afterglow}\label{sec:x-aft}

\begin{figure}
\includegraphics[width=9cm,angle=0]{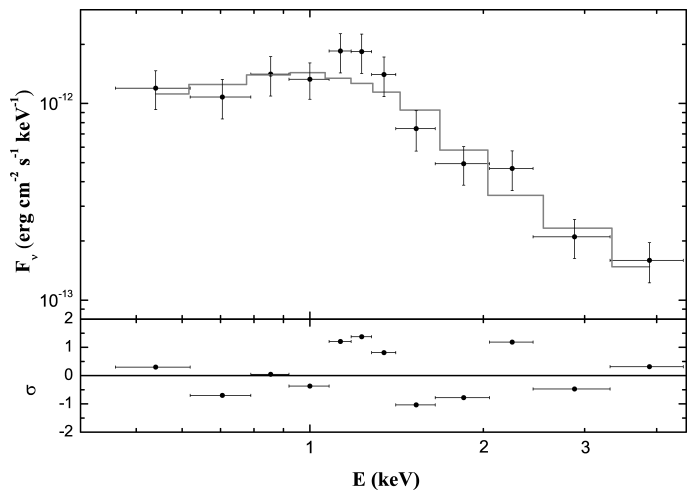}
\caption{X-ray spectrum of the afterglow of GRB 080514B obtained in the
        photon counting mode between 0.43 and 0.57 days after the
        trigger. The spectrum was fitted with an absorbed power-law. The
        lower panel shows the residuals of the best fit.}
\label{fig:sedX}
\end{figure}

\begin{figure}
\includegraphics[width=9cm]{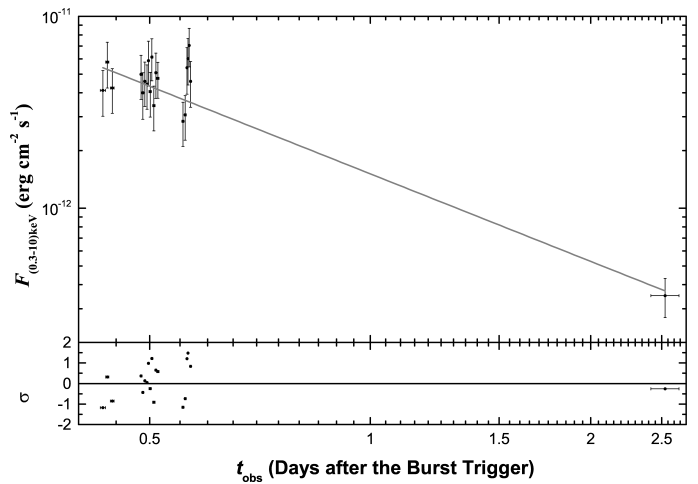}
\caption{X-ray light curve of the afterglow of GRB 080514B observed with
        \swift XRT. The observation started 
        0.43 days and finished about 2.6 days after the trigger. To create 
        the flux-calibrated light curve, the spectral fit was used to derive
        an energy conversion factor of $6.1\,\times\,10^{-11}\, \rm{erg}
        \,\rm{cm}^{-2}\, \rm{counts}^{-1}$.}
\label{fig:xlc}
\end{figure}

Fitting the spectrum of the first observing block ($0.43 - 0.57$ days; total
exposure time 5916~s) with an absorbed power-law\footnote{The T\"ubingen model 
was used to take in account the absorption, see Wilms et al. (\cite{Wilms}).} 
results in a spectral slope\footnote{For the  flux density of the afterglow we 
use the usual convention $F_\nu (t) \propto t^{-\alpha} \nu^{-\beta}$.} of 
$\beta_{\rm X} = 1.01 ^{+0.28} _{-0.25}$ and an effective hydrogen column density 
of $N_{\rm H} =  1.4 ^{+0.9} _{-0.8}\times 10^{21} \rm{cm}^{-2}$ ($\chi^2 /\rm{d.o.f.} 
= 7.97/9$), in agreement with values reported by Page et al. (2008; Fig.~\ref{fig:sedX})
($1\sigma$ uncertainties). We were unable to constrain the possibility of spectral evolution.
The derived hydrogen column density is higher than the 
Galactic value of $N_{\rm H} = 0.375\, \times\,10^{21}\ \rm{cm}^{-2}$ based on radio 
observations (Kalberla et al. 2005). This suggests the presence of additional absorption by gas
in the host galaxy. 

The canonical X-ray afterglow light curve derived by Nousek et al. (2006) shows a
transition from a plateau to the normal decay phase between $0.1$ and $1$ days
post-burst and a jet break thereafter. Unfortunately, for GRB 080514B 
at early times (0.463 to 0.694 days) the X-ray light curve exhibits substantial scatter, as it has also been 
the case for other X-ray afterglows (cf. O'Brien et al. 2006). This, together with
the lack of data thereafter, makes it impossible to decide whether there was a
plateau phase at early times (0.463 to 0.694 days), a flare, or a break in the
decay between 0.694 to 2.315 days. Assuming a simple power-law decay, the light
curve is well described by a temporal decay index of $\alpha_{\rm X}=
1.52\pm0.14$ ($\chi^2$/d.o.f. = 17.68/18). A smoothly broken power-law is
statistically not favoured (Fig.~\ref{fig:xlc}).

\subsection{The optical afterglow}\label{sec:optaft}

\begin{figure}
\includegraphics[width=9cm]{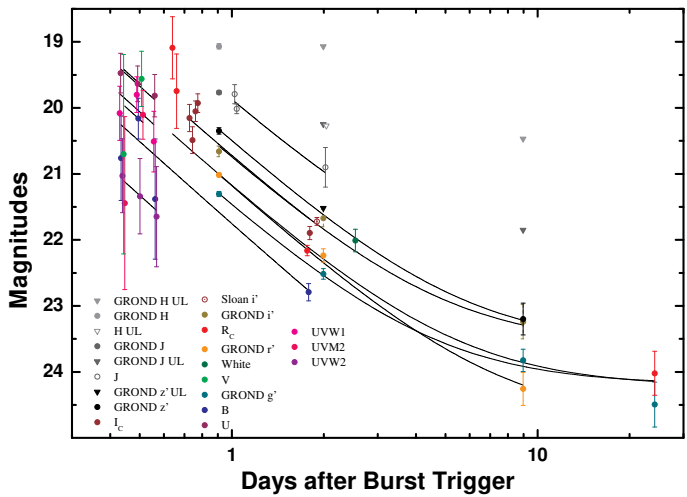}
\caption{
Data of the optical/NIR afterglow of GRB 080514B. The afterglow is  detected
in all observed bands ($uvw2$ to $H$) except for the $K$ band. Upper limits
are given as downward pointing triangles. The lines show the simultaneous
fit with a single power-law plus host galaxy component. The decay slope is 
$\alpha_{\rm opt}=1.67\pm0.07$.}
\label{fig:optlc}
\end{figure}

While the optical afterglow is detected over a broad range of filters, from
the \swift UVOT $uvw2$ band to the $H$ band (160-1700 nm), the data set is sparse, with some
scatter (Fig.~\ref{fig:optlc}). To determine the slope of the light curve decay
as well as the spectral energy distribution (SED) of the afterglow,
 we simultaneously fit all 14 bands with detections 
(excluding the UVOT White filter measurement) with a single
power-law and an added host galaxy component for those bands where the late
flattening indicates that the afterglow has become fainter than the host. From
this fit ($\chi^2/$d.o.f. $=1.51/25$), we find a decay
slope $\alpha_{\rm opt}=1.67\pm0.07$. Unfortunately, this value alone is insufficient to
decide whether this is a pre-break or a post-break decay. Light curves with such 
a (steep) pre-jet break decay slope or with such a (flat) post-jet break decay 
slope have both been observed (see, e.g., Zeh et al. 2006, Kann et al. \cite{Kann2008} for 
compilations of optical afterglow data). We thus find no evidence for a jet break.

\subsection{SED and photometric redshift}\label{sec:sedz}

\begin{figure}
\includegraphics[width=9cm]{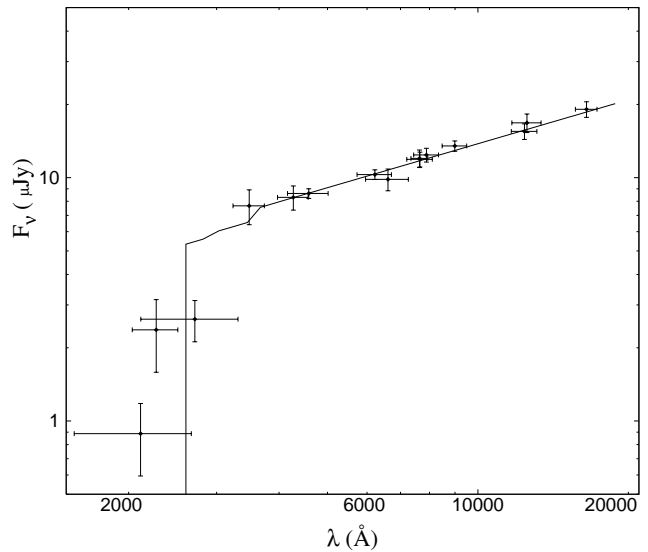}
\caption{
The SED of the optical/NIR afterglow at 1 day after the burst fitted using the \emph{HyperZ} tool. From
the left the observed flux density in the filters: UVOT $uvw2$, $uvm2$, $uvw1$, $U$ and
$B$, GROND $g^\prime$, GROND $r^\prime$, $R_C$, GROND $i^\prime$, Gemini $i'$, $I_C$, GROND $z^\prime$, GROND
$J$, NEWFIRM $J$ (see Tab. \ref{mags}) and GROND $H$. The widths of the bands correspond to their approximation by a 
Gaussian filter (see Bolzonella et al. 2000).}
\label{fig:sedfit}
\end{figure}

The simultaneous fitting procedure described in \S \ref{sec:optaft} yields magnitudes 
normalized to one day after the GRB for each band, which define the spectral energy 
distribution (SED) of the afterglow. We find no evidence for strong chromatic evolution 
but caution again that the data are sparse and often have low signal-to-noise values. The 
SED is well described by a simple power-law with spectral slope $\beta_{\rm opt}=0.64 \pm 0.03$ 
($\chi^2 /\rm{d.o.f.} = 8.58/10$) from the $H$ band to the $U$ band (Fig.~\ref{fig:sedfit}). 
 The UVOT $V$ band is an outlier of the fit but it does not significantly disturb the result. 
We do not find evidence for dust in the host galaxy, which would create spectral 
curvature. The three UVOT UV filters, on the other hand, show a much steeper slope. Such a 
feature cannot be explained by standard models of dust extinction, but it is the signature of 
the Lyman dropout effect due to intergalactic hydrogen along the line of sight to a relatively 
large redshift. Using \emph{HyperZ} (Bolzonella et al. 2000), and assuming $A_{\rm V}^{\rm host}$=0, 
the best fit indcates a photometric redshift of $z=1.8^{+0.4}_{-0.3}$ ($1\sigma$ 
uncertainties, see Avni \cite{Avni}), in agreement with the 
constraint of $z<2.3$ based on Gemini-North observations (Perley et al. 2008) and the pseudo 
redshift of $z=1.76\pm0.30$ based on the burst spectrum (Pelangeon \& Atteia 2008). On the other 
hand, it is intermediate between the two redshift estimations presented by Gendre et al. (2008).
Excluding the $uvw2$ filter from the fit, does not change the obtained photometric redshift. 
Also, doubling the assumed Galactic extinction value to $A_V=0.38$, 
does not change the deduced photometric redshift notebly. However, 
the shape of the SED then indicates that we have overcorrected for the extinction.

Fixing $z=1.8$, we re-fit the SED (now excluding the UVOT UV filters) with dust models for the 
Milky Way, Large and Small Magellanic Clouds (see Kann et al. 2006 for the procedure). 
In all cases, adding $A_V$ as an additional parameter does not improve the fits significantly, 
and the derived extinction is zero within errors for all three cases too (at 3$\sigma$ confidence, 
$A_V\leq0.06$ for MW, $\leq0.17$ for LMC and $\leq0.14$ for SMC dust). No evidence for a 2175 {\AA} 
feature (which would lie close to the $R_C$ and $r^\prime$ bands) is apparent, and no discrimination 
is possible between dust models.

The assumption of zero extinction is consistent with several studies (Starling et al. 2007; 
Schady et al. 2007) on the dust-to-gas ratios in host galaxies of GRBs where it is shown 
that the observed GRBs occur in low-metallicity environments. 

For an assumed redshift of $z=1.8$, and fixing the 
Galactic hydrogen column density to the value  given by Kalberla et al. (2005), this implies a 
host-intrinsic column density of $ N^{\rm host} _{\rm H} = 8.7 ^{+9.0} _{-7.3}\times 10^{21}  
\rm{cm}^{-2}$ and an unabsorbed spectral index of $\beta_{\rm X} = 0.94 ^{+0.24} _{-0.21}$ 
($\chi^2 /\rm{d.o.f.} = 8.12/9$). The deduced value for the spectral slope is consistent 
with the mean value found for \emph{Swift} X-ray afterglows (O'Brien et al. 2006).

Using the derived spectral slope and redshift, 
the absolute magnitude of the afterglow is $M_B=-22.17\pm0.2$ and $M_B=-20.17\pm0.5$, at one and four 
days after the GRB, respectively (for the method see Kann et al. 2006, 2008; no extinction is assumed).
A comparison with the sample presented in Kann et al. (2008) shows that these are typical 
values for a GRB afterglow, i.e., GRB 080514B is neither exceptionally bright or faint.


\begin{table}
\centering
\begin{tabular}{llrlr}
\toprule
afterglow model				& \multicolumn{2}{c}{Optical}		& \multicolumn{2}{c}{X-ray}		\\
				&predicted $\alpha$	& $\sigma$-level& predicted $\alpha$	& $\sigma$-level\\
\midrule
\multicolumn{5}{c}{iso}\\
ISM, wind, $\nu_c<\nu$		&$0.46\pm0.04	$	&$-14.82$	&$1.02^{+0.42} _{-0.38}$&$-1.26$		\\
ISM, $\nu<\nu_c$		&$0.96\pm0.04$		&$-8.70$	&$1.34^{+0.42} _{-0.38}$&$-0.01$		\\
wind, $\nu<\nu_c$		&$1.46\pm0.04	$	&$-2.57	$	&$2.02^{+0.42} _{-0.38}$&$1.24$		\\
\midrule
\multicolumn{5}{c}{jet}\\
ISM, wind, $\nu_c<\nu$		&$1.28\pm0.06	$	&$-4.35$	&$2.02^{+0.56} _{-0.50}$&$0.96$		\\
ISM, wind, $\nu<\nu_c$		&$2.28\pm0.06	$	&$6.80$		&$3.02^{+0.56} _{-0.50}$&$2.89$		\\
\bottomrule
\end{tabular}
\caption{Predicted values for the temporal slopes $\alpha$ for various
afterglow scenarios based on the measured spectral slope $\beta$ in the optical/NIR 
bands (\ref{sec:sedz}) and in the X-ray band (\ref{sec:x-aft}). Assuming a relativistic jetted explosion, for
observations at $t<t_{\rm break}$ (pre-break time) the isotropic model holds,
whereas for $t>t_{\rm break}$ (post-break time) the jet model applies (e.g.,
Sari et al. 1999). The $\sigma$-level is the difference of the predicted and
the observed temporal slope, normalized to the square root of the sum of their
quadratic errors.}
\label{table:alphabeta}
\end{table}

\subsection{The host galaxy}

A galaxy underlying the position of the optical transient is detected in all
GROND optical bands at 8.9 days, as well as in the deep Keck $g$ and $R$-band
images obtained 24.13 days post-burst. Using the stacked GROND $g'r'i'z'$
images, the  coordinates of this galaxy are R.A., Dec. (J2000) = $21^{\rm
h}31^{\rm m}22.^{\rm s}68$, $+00^\circ42'28\farcs8$, which is
$0\farcs3\pm0\farcs2$ off from the position of the optical afterglow. Assuming a
cosmological model with $H_0=71$ km s$^{-1}$ Mpc$^{-1}, \Omega_{\rm M}= 0.27,
\Omega_\Lambda=0.73$ (Spergel et al. \cite{Spergel2003}), for a redshift of
1.8 the offset of the optical transient from the center of its likely host galaxy
is $2.6\pm1.7$ kpc. 

Assuming a power-law spectrum for the putative host galaxy of the form $F_\nu \propto
\nu^{-\beta_{\rm gal}}$, its absolute $R_C$-band magnitude is $M_R = m_R - \mu
- k$, where $\mu=45.70$ mag is the distance modulus and $k$ is the
cosmological $k$-correction, $k= -2.5 (1-\beta_{\rm gal}) \log (1+z$). Hence,
$M_R = -21.53 + 1.12 (1-\beta_{\rm gal}) \pm 0.3$, which for $\beta_{\rm
gal}$=0.45, as it follows from the third epoch GROND $g'r'i'z'$ data, 
makes this galaxy approximately 0.5 mag more luminous than the
 characteristic magnitude of the Schechter function describing the $r$-band 
luminosity function of galaxies in the Las Campanas redshift survey (Lin et al. 1996). 
Its $R$-band magnitude matches well into the distribution of host magnitudes 
of long bursts at this redshift (Guziy et al. 2005; Savaglio et al. 2008).

As mentioned in \S \ref{sec:x-aft} and \ref{sec:optaft}, based on the light curve alone 
we cannot unambiguously decide whether the data belong to the pre-jet break phase or to
the post-jet break phase. However, using the standard 
$\alpha-\beta$ relations (e.g., Sari et al. 1999), the data favors an
isotropic model and a wind environment with, at $t$=1 day, the position of
the cooling frequency in between the optical/NIR and the X-ray band
(Tab.~\ref{table:alphabeta}). The difference of the spectral slopes between both bands is
$\Delta\beta=0.35^{+0.25}_{-0.22}$, which has to be compared to the theoretical value of $\Delta\beta=0.5$.
Unfortunately, the non-detection of the afterglow at 86~GHz does not constrain the shape of the SED further.

\section{Summary and conclusions}

To our knowledge, GRB 080514B is the first gamma-ray burst detected above 30
MeV for which an afterglow has been found in the X-ray band and in the
optical/NIR bands. Based on our ground-based follow-up observing campaign, in
combination with \emph{Swift} UVOT and \emph{Swift} XRT data starting 0.4 days
after the burst, we find: (1) The X-ray/optical/NIR light curve after 0.4 days
is well described by a single power-law with no sign of a jet break. 
(2) The SED of the afterglow indicates strong Lyman blanketing at short wavelengths, 
implying a photometric redshift of $z=1.8^{+0.4}_{-0.3}$. This is the first redshift determination 
for a GRB with prompt emission detected beyond 30 MeV. We find no evidence for 
extinction by dust in the GRB host galaxy. (3) A comparison of the observed light 
curve decay with the spectral energy distribution favours a model in which the afterglow 
blast wave propagated in a wind medium. (4) The intrinsic properties of the optical 
afterglow are typical for other long-duration GRBs. (5) The putative GRB host galaxy, 
identified in our GROND and Keck images, has $R_c=24.2 \pm 0.3$ and an absolute $R$-band
magnitude of $M_R=-20.9 \pm 0.3$. (6) The optical transient was offset from the center 
of its host by $2.6 \pm 1.7$ kpc. 

We conclude that according to our data set the afterglow properties
as well as the properties of the host galaxy match into what is known about the corresponding 
properties of the long burst sample. The only property that make this burst remarkable is its 
detection above 30 MeV.

\begin{acknowledgements}
A.R., P.F., D.A.K. and S.K. acknowledge support by DFG Kl 766/11-3 
and 13-1. R.F. and S.S. were supported by the Th\"uringer 
Landessternwarte Tautenburg. 
T.K acknowledges support by the DFG cluster 
of excellence 'Origin and Structure of the Universe'.
J.P.U.F. acknowledges support by the DNRF and 
J. Gorosabel by the programmes ESP2005-07714-C03-03 and AYA2007-63677.
We acknowledge D. Malesani for a careful reading of the manuscript, 
P. E. Nissen and W. J. Schuster for performing the NOT observations as well as Caroline Pereira, A. Pimienta,
 E. Curras and C. Pereira for performing the IAC80 observations.
This work made use of data supplied by the UK Swift Science Data Centre
at the University of Leicester.
\end{acknowledgements}


\end{document}